\documentclass[10pt,twocolumn,letterpaper,superscriptaddress]{revtex4}

\usepackage[paperwidth=215mm,paperheight=300mm,centering,hmargin=1.5cm,vmargin=1.5cm]{geometry}
\usepackage{multibbl}
\usepackage[pdftex]{graphicx}
\usepackage{amsmath,SIunits}
\usepackage[latin1]{inputenc}
\usepackage{latexsym,stmaryrd}
\usepackage[sort&compress]{natbib}
\usepackage[pdftex]{graphicx}
\usepackage{ulem}
\usepackage{color}
\bibpunct{[}{]}{,}{n}{}{,}

\makeatletter
\newcommand*{\citenst}[2][]{%
  \begingroup
  \let\NAT@mbox=\mbox
  \let\@cite\NAT@citenum
  \let\NAT@space\NAT@spacechar
  \let\NAT@super@kern\relax
  \renewcommand\NAT@open{[}%
  \renewcommand\NAT@close{]}%
  \citep{#2}%
  \endgroup
}
\makeatother

\renewcommand{\figurename}{\textbf{Figure}}

\begin{document}

\title{Engineering nanoscale hypersonic phonon transport}

\author{O. Florez}
\email{omar.florez@icn2.cat}
\affiliation{Catalan Institute of Nanoscience and Nanotechnology (ICN2), CSIC and BIST, Campus UAB, Bellaterra, 08193 Barcelona, Spain}
\affiliation{Dept. de F\'{i}sica, Universitat Autonoma de Barcelona, 08193 Bellaterra, Spain}
\author{G. Arregui}
\affiliation{Catalan Institute of Nanoscience and Nanotechnology (ICN2), CSIC and BIST, Campus UAB, Bellaterra, 08193 Barcelona, Spain}
\altaffiliation[current address:]{Department of Photonics Engineering, DTU Fotonik, Technical University of Denmark, Building 343, DK-2800 Kgs. Lyngby, Denmark}
\author{M. Albrechtsen}
\affiliation{Department of Photonics Engineering, DTU Fotonik, Technical University of Denmark, Building 343, DK-2800 Kgs. Lyngby, Denmark}
\author{R. C. Ng}
\affiliation{Catalan Institute of Nanoscience and Nanotechnology (ICN2), CSIC and BIST, Campus UAB, Bellaterra, 08193 Barcelona, Spain}
\author{J. Gomis-Bresco}
\affiliation{Catalan Institute of Nanoscience and Nanotechnology (ICN2), CSIC and BIST, Campus UAB, Bellaterra, 08193 Barcelona, Spain}
\altaffiliation[current address:]{Departament de F\'{i}sica Aplicada, Universitat de Barcelona, IN2UB, 08028 Barcelona, Spain}
\author{S. Stobbe}
\affiliation{Department of Photonics Engineering, DTU Fotonik, Technical University of Denmark, Building 343, DK-2800 Kgs. Lyngby, Denmark}
\affiliation{NanoPhoton - Center for Nanophotonics, Technical University of Denmark, {\O}rsteds Plads 345A, DK-2800 Kgs.\ Lyngby, Denmark}
\author{C. M. Sotomayor-Torres}
\affiliation{Catalan Institute of Nanoscience and Nanotechnology (ICN2), CSIC and BIST, Campus UAB, Bellaterra, 08193 Barcelona, Spain}
\affiliation{ICREA - Instituci\'o Catalana de Recerca i Estudis Avan\c{c}ats, 08010 Barcelona, Spain}
\author{P. D. Garc\'{i}a}
\email{pd.garcia@csic.es}
\altaffiliation[current address: ]{Instituto de Ciencia de Materiales de Madrid (ICMM), Consejo Superior de Investigaciones Cient\'{i}ficas (CSIC), Calle Sor Juana In\'{e}s de la Cruz, 3, 28049 Madrid, Spain.}
\affiliation{Catalan Institute of Nanoscience and Nanotechnology (ICN2), CSIC and BIST, Campus UAB, Bellaterra, 08193 Barcelona, Spain}

\date{\today}

\small

\begin{abstract}
Controlling the vibrations in solids is crucial to tailor their mechanical properties and their interaction with light.\ Thermal vibrations represent a source of noise and dephasing for many physical processes at the quantum level.\ One strategy to avoid these vibrations is to structure a solid such that it possesses a phononic stop band, i.e., a frequency range over which there are no available mechanical modes.\ Here, we demonstrate the complete absence of mechanical vibrations at room temperature over a broad spectral window, with a 5.3 GHz wide band gap centered at 8.4 GHz in a patterned silicon nanostructure membrane measured using Brillouin light scattering spectroscopy.\ By constructing a line-defect waveguide, we directly measure GHz localized modes at room temperature.\ Our experimental results of thermally excited guided mechanical modes at GHz frequencies provides an efficient platform for photon-phonon integration with applications in optomechanics and signal processing transduction.

\end{abstract}

% insert suggested PACS numbers in braces on next line
 \pacs{(42.25.Dd, 62.25.-g, 46.65.+g, 42.50.Wk)}

\maketitle

Nanostructured materials offer the possibility to manipulate the mechanical vibrations of a solid over a specified spectral bandwidth.\ This in turn enables the control of light-matter interactions in the visible and near-infrared regimes for optomechanical applications ranging from high-resolution accelerometers \cite{accelerometer} to mass and force sensors \cite{masssensor, forcesensor}, in addition to providing fundamental insight into phenomena such as quantum ground-state cooling \cite{sidebandcool, lasercool}.\ By periodically distributing the mass within a system, it is possible to engineer its mechanical modes \cite{sigalas, kushawa} and open frequency windows over which the destructive interference of scattered waves forbids any phonon propagation \cite{martinez, gorishnyy}.\ This approach enables engineering of the thermal conductance of the structure \cite{zen} and allows for the routing of phonons at the mesoscale \cite{eichenfield, rouhani}.\ Although full-gap GHz phononic crystals are widely used in optomechanical systems to create phononic shields \cite{maccabe}, waveguides \cite{kfang, rnpatel}, and cavities \cite{snowflake2, jordinc}, clear and direct experimental evidence of a complete omnidirectional phononic band gap at hypersonic (GHz) frequencies is still lacking.\ Existing experimental work is generally limited to MHz frequencies, using piezoelectric materials to drive the system \cite{mohammadi, soliman, gorisse}, requiring varying interdigitated electrodes to probe different frequencies and propagation directions. In the GHz regime, only partial and narrow mechanical band gaps (with up to $8{\%}$ gap to mid-gap ratio) have been shown using assembled platforms such as colloidal crystals \cite{cheng} or two-dimensional phononic crystal membranes \cite{bart1}.\ Furthermore, the control and guiding of mechanical waves at GHz frequencies has been difficult to achieve or measure, relying on complex optomechanical systems \cite{kfang, rnpatel}, or nonlinear stimulated phenomena \cite{moli}.

\begin{figure*}
\includegraphics[width=2\columnwidth]{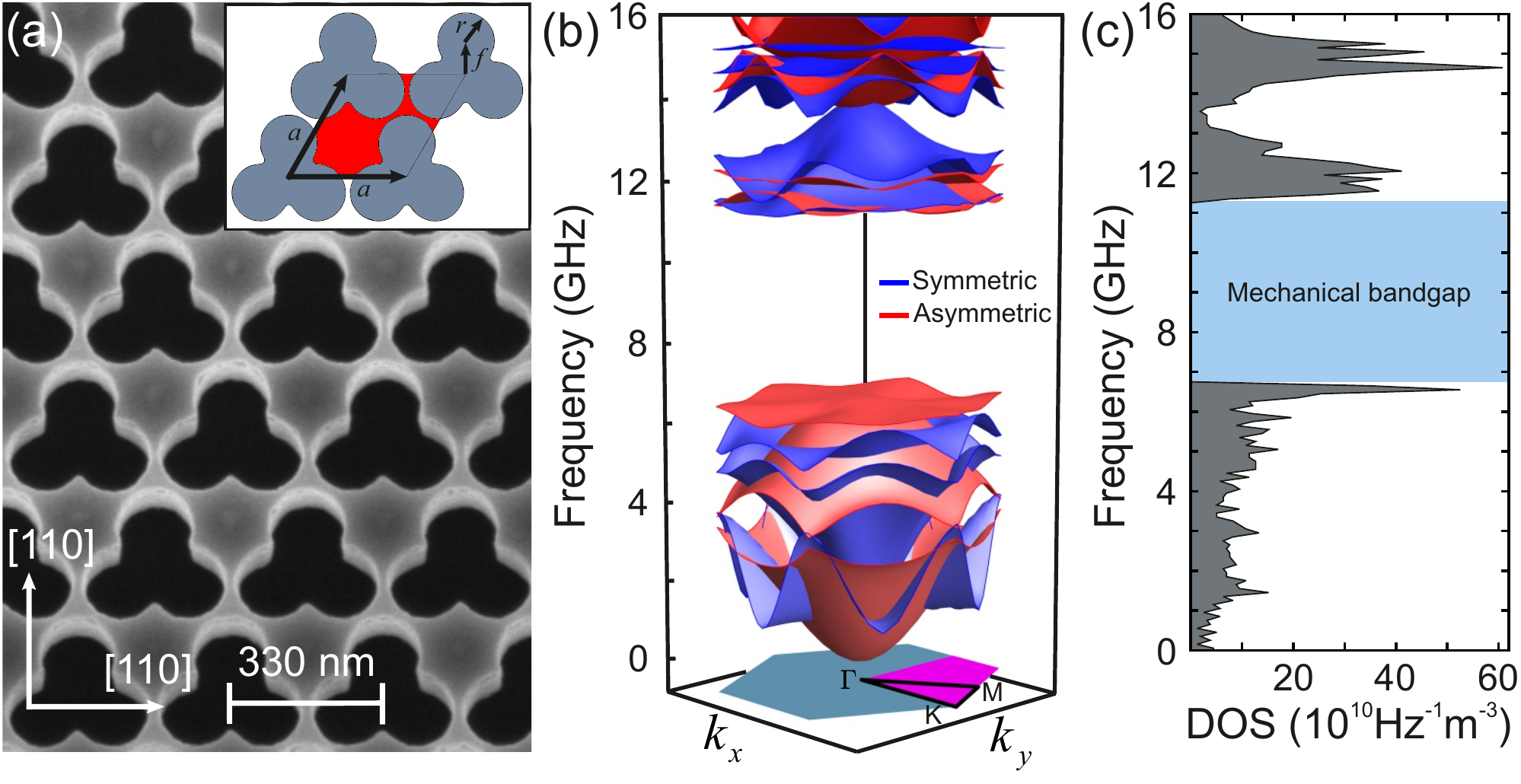}
\caption{ \label{1} \textbf{Shamrock phononic insulator.} (a) Scanning-electron micrograph (tilted-top view) of the fabricated structure on a silicon-on-insulator substrate with a thickness of $t$ = 220 nm.\ The inset schematically illustrates the geometrical parameters of the unit cell (highlighted in red) with lattice constant $a$ = 330 nm, hole radius $r = 0.22a$, and the distance between the center of the shamrock and the center of each circle $f = 2r/\sqrt{3}$.\ (b) Simulated 3D mechanical dispersion relation of the crystal over the first Brillouin zone. Blue and red curves indicate the symmetric and asymmetric modes with respect to the middle plane of the silicon slab at $t/2$.\ (c) Calculated phononic density of states (DOS) of the structure.\ The light-blue region highlights the full mechanical gap spanning 6.7 GHz to 11.4 GHz.}
\end{figure*}

Here, we report direct experimental evidence of a wide full phononic gap with a central frequency at 8.4 GHz and a spectral width of 5.3 GHz (a gap to mid-gap ratio of 64\%) in a free-standing patterned silicon membrane phononic crystal.\ Additionally, we create a line-defect waveguide with the same geometry in which we directly measure two guided modes at 5.7 and 7.1 GHz within the band gap at room temperature.\ We demonstrate the spectral tunability of the mechanical gap from approximately 4 to 12 GHz, which subsequently enables spectral tunability of the guided modes.\ All the structures measured here are fabricated on a silicon-on-insulator (SOI) platform, which readily enables integration with electronic and photonic circuits.\ Figure \ref{1}(a) shows a scanning electron micrograph (SEM) of the fabricated pattern composed of a triangular array of ``shamrocks'' \cite{immo, arreguiprb}, formed by three tangential circles with nominal parameters of thickness $t=220$ nm, period $a=330$ nm, and radius $r=0.22a$, as detailed in the inset of Fig. \ref{1}(a) and Fig. S1 in the supplementary information (SI).\ We calculate the mechanical dispersion relation of the structure by solving the full three-dimensional elastic wave equation using finite-element (FEM) simulations performed with COMSOL Multiphysics \cite{comsol}.\ Figure \ref{1}(b) plots the symmetric (blue) and asymmetric (red) acoustic modes with respect to the mid-plane of the silicon slab, calculated over the entire first Brillouin zone (BZ) of the crystal.\ We use the geometrical parameters extracted from SEM images to more accurately simulate the real shape of the fabricated crystal (see Fig. S3 in SI for a statistical analysis).\ We also take into account the anisotropy of the silicon stiffness tensor and its particular orientation with respect to the fabricated samples, as detailed in Fig. \ref{1}(a). Due to this mechanical anisotropy, the irreducible BZ is determined by the first quarter of the hexagon highlighted on the bottom part of Fig. \ref{1}(b) (see section S3 in SI).\ A full mechanical gap opens between the 6th and 7th bands, from 6.7 GHz up to 11.4 GHz (gap to mid-gap ratio of 52 \%) which results in the complete depletion of the phonon density of states (DOS) over this frequency range, as shown in Fig. \ref{1}(c).\ The particular shape of the shamrock crystal, which is comprised of large masses connected by small necks, enables a distribution of the mass within the unit cell that results in this broad mechanical gap.\ A direct link exists between the spectral width of the gap and the narrow necks (shorter distance between shamrocks): a larger radius leads to narrower connected neck regions, which subsequently widens the gap \cite{snowflake0}.

\begin{figure*}
\includegraphics[width=2\columnwidth]{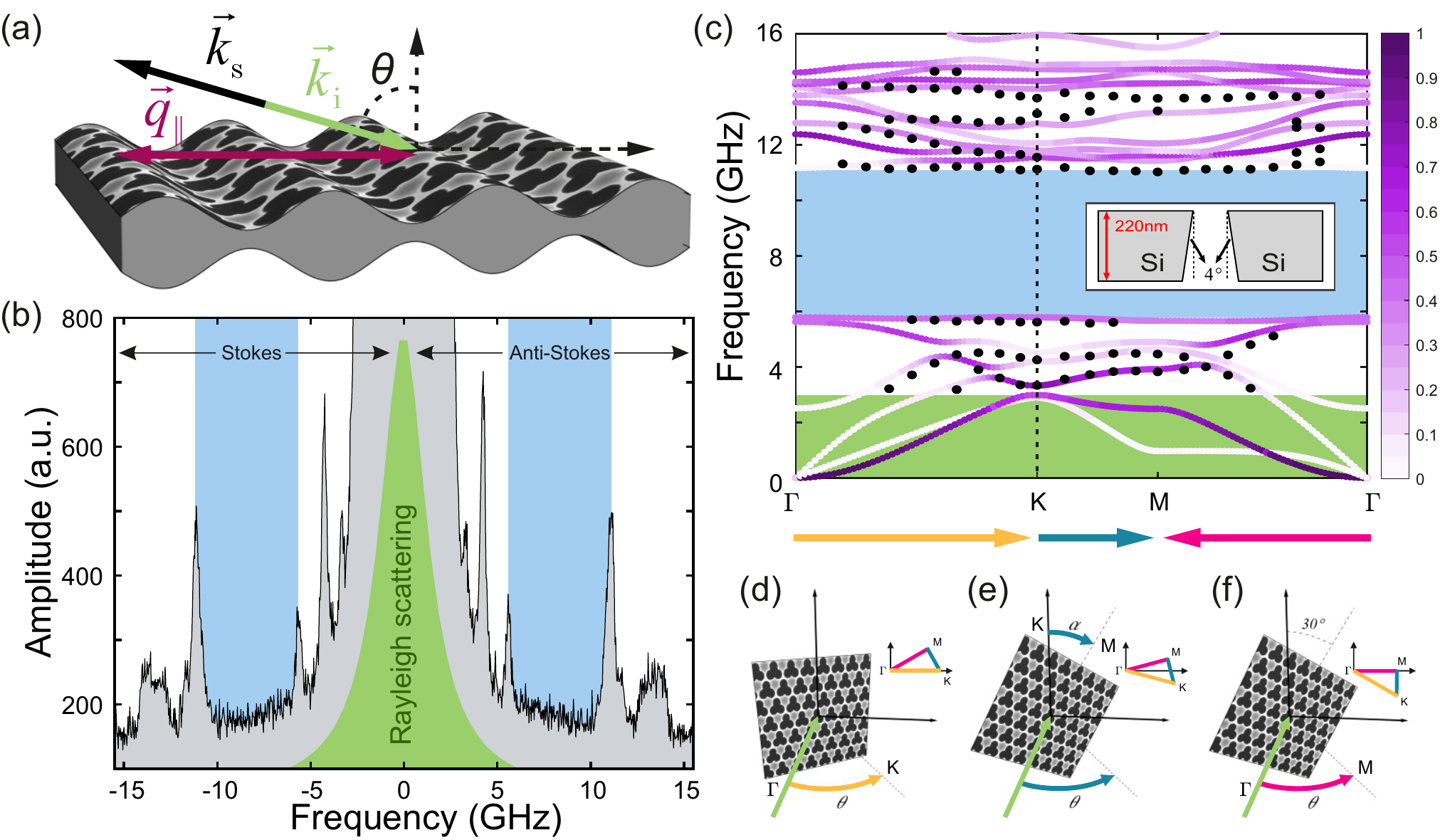}
\caption{ \label{2} \textbf{ Brillouin light scattering spectroscopy.} (a) Schematic illustration of Brillouin scattering with the phase-matching condition for the backward configuration used in the experiments.\ Here, $\vec{k}_i$ and $\vec{k}_s$ represent the incident and the scattered light, respectively, and $\vec{q}_{\parallel}$ is the parallel mechanic wavevector.\ The magnitude of $\vec{q}_{\parallel}$ depends on the incident angle, where $q_{\parallel}=2k_i\sin{\theta}$.\ (b) Measured Brillouin scattering spectrum for an incident angle of $\theta=32.5^\circ$ with \textit{p}-polarized light.\ The green central peak stems from elastic Rayleigh scattering.\ Negative and positive frequency peaks on either side of this large central peak correspond to Stokes and anti-Stokes contributions, respectively.\ The light-blue regions highlight the mechanical gap.\ (c) Calculated dispersion relation based on the geometrical parameters obtained from SEM images of the fabricated samples that include a $4^\circ$ sidewall angle correction in the vertical profile (inset).\ The black dots represent the measured frequencies of vibrational modes for different angles and the vertical dotted line indicates the frequencies obtained from the measured spectrum shown in (b).\ The intensity color scale represents the normalized coupling coefficients for the MB perturbation.\ (d), (e) and (f) indicate the direction in which the sample is physically rotated to scan along the highest-symmetry directions $\Gamma K$, $KM$, and $\Gamma M$, respectively.\ The green arrows indicate the direction of the incident laser light while the other colored arrows correspond to the rotation direction during measurements, which represent (and are color-consistent with) the highest-symmetry direction indicated in (c).}
\end{figure*}

We use Brillouin light-scattering spectroscopy~\cite{abalandin, carlotti} to reconstruct the mechanical dispersion relation of the system.\ For simplicity, we probe the band structure along the $\Gamma KM\Gamma$ path, highlighted at the bottom of Fig. \ref{1}(b), as the edges of the gap do not change in frequency with respect to the irreducible BZ (see section S2 in SI).\ When incident light with frequency $\nu_i$ and wavevector $\vec{k}_i$ reaches the surface of the sample with a certain angle, $\theta$, as illustrated in Fig. \ref{2}(a), part of it is linearly scattered while another small part is nonlinearly scattered in all directions by thermally activated acoustic phonons.\ This scattering process occurs either by the elasto-optic \cite{boyd} or the moving-boundary (MB) \cite{johnson} mechanism.\ The former is a volumetric effect caused by the acoustic modulation of the dielectric constant $\epsilon$ inside the material, while the latter is a surface effect induced by the movement of phonons that creates corrugation at the interface.\ The interplay between these two effects can result in the enhancement \cite{vanlaer} or the cancellation \cite{selfcancel} of the scattering process.\ Given the high refractive-index contrast between silicon and the surrounding air, and the small volume of interaction, determined by the direction of the incident beam and the thickness of the suspended structure, the scattering process here is dominated by the MB mechanism (see S4 in the SI).\ Our experiment collects the backscattered signal, $\vec{k}_s$.\ For this configuration, the phase-matching condition for the mechanical wavevector $\vec{q}_{\parallel}$, which lies parallel to the surface, is determined by
\begin{equation}
\label{pmc}
q_{\parallel}=2k_{i}\sin\theta=\frac{4\pi}{\lambda_i}\sin\theta,
\end{equation}
where $k_{i}=2\pi/\lambda_i$.\ Therefore, it is possible to probe different mechanical wavevectors by changing the angle of incidence of light $\theta$, illustrated in Fig. \ref{2}(a), and subsequently map the dispersion relation of the acoustic phonons.\ All measurements were taken by focusing a green laser ($\lambda_i$ = 532 nm) that is \textit{p}-polarized with respect to the sagittal plane formed by the angle $\theta$ (see Fig. S6 in SI). The scattered light that was analyzed is also the \textit{p}-polarized component.\ Although in-plane and out-of-plane mechanical modes in bulk materials and membranes can be selectively detected using light polarization~\cite{jcuffe}, mechanical modes in phononic crystals are generally mixed.\ Therefore, we do not obtain different information by considering different polarizations of incident and analyzed light.
\begin{figure*}
\includegraphics[width=2\columnwidth]{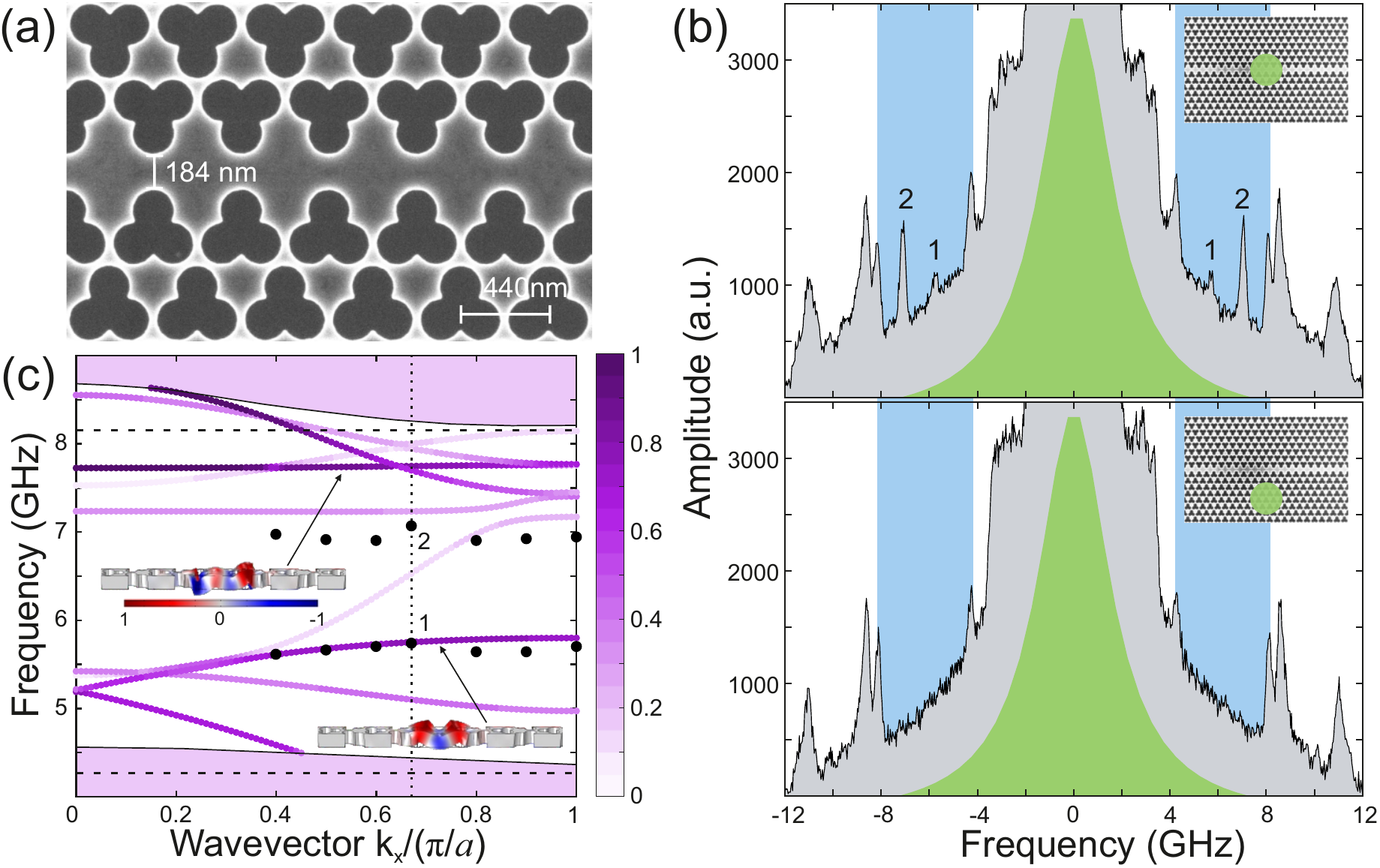}
\caption{ \label{3} \textbf{Hypersonic phononic waveguide} (a) SEM image of a shamrock phononic waveguide with a lattice period of $a$ = 440 nm and waveguide width of $w$ = 184 nm.\ The thickness and radius of the structure are the same as in previous structures ($t=$ 220 nm, $r = 0.22a$).\ (b) Measured Brillouin scattering spectra in the waveguide (top) and surrounding phononic crystal (bottom) as is illustrated in the insets, for an incident light angle of $23.8^\circ$.\ The spectral width of the measured gap is indicated by the blue regions.\ Two peaks whose frequencies correspond with the guided modes of the system appear inside the gap in the spectrum of the waveguide (top).\ (c) Calculated dispersion relation of the waveguide.\ The intensity color scale represents the normalized coupling coefficient for the MB perturbation.\ The horizontal and vertical dotted lines indicate the mechanical band edges and phononic wave vector, respectively, while the black dots represent the frequencies of peaks 1 and 2, all for the top waveguide spectrum shown in (b).\ The insets in (c) display the mode profiles for the indicated bands where the color represents the normalized out-of-plane displacement.}
\end{figure*}
Figure \ref{2}(b) plots the mechanical spectrum measured with incident angle $\theta=32.5^\circ$, which corresponds to the high-symmetry point $K$ in reciprocal space.\ The central peak highlighted in green corresponds to the elastic (Rayleigh) scattered signal.\ Positive and negative frequencies correspond with anti-Stokes and Stokes contributions respectively, which are equally likely in a stochastic process such as spontaneous Brillouin scattering \cite{boyd, brillouin}.\ All the peaks observed in the spectrum correspond to vibrational modes of the system and their amplitudes depend on the scattering efficiency of each mode with the incident laser light \cite{loudon}, which is proportional to the displacement of the boundaries, as previously mentioned and detailed in section S4 of the SI.\ We obtain the phonon frequencies by fitting each of the observed peaks to Lorentzian line shapes and extracting the mean value between the resonant frequencies of the Stokes and anti-Stokes components.\ Figure \ref{2}(c) plots the mechanical dispersion relation along the $\Gamma K M \Gamma$ path.\ The intensity color scale represents the normalized coupling coefficients for the MB perturbation.

The sidewalls of our structures are angled at $4^\circ$ relative to vertical, which is taken into account in our band structure calculation, as shown in the inset of Fig. \ref{2}(c).\ This breaks the up-down symmetry and the mechanical modes of the real structure therefore can not be classified by their symmetry with respect to the mid plane of the slab as done previously in Fig. \ref{1}(b).\ For this reason, all bands in Fig. \ref{2}(c) are indicated with the same color and only change in intensity to indicate the scattering efficiency of each mode and wavevector.\ We observe that this small correction to the vertical profile causes a displacement of about 1 GHz in the bands below the band gap which becomes evident upon comparing Fig. \ref{2}(c) and Fig. \ref{1}(c).\ Additionally, the gap to mid-gap ratio increases from 52 \% in Fig. \ref{1}(b) and \ref{1}(c), to 64 \% in Fig. \ref{2}(c).\ The black dots in Fig. \ref{2}(c) are the measured frequencies of the peaks as the incident angle is varied. The vertical dotted line indicates the position of frequencies obtained for the spectrum shown in Fig. \ref{2}(b).\

To resolve the full mechanical dispersion relation, we map the highest-symmetry directions of the Brillouin zone: $\Gamma K$, $KM$, and $\Gamma M$. The $\Gamma K$ path is measured by varying the angle of incidence $\theta$ from zero to $32.5^\circ$, as depicted in Fig. \ref{2}(d) where the green arrow represents the incident laser.\ Here, the value of the maximum angle $\theta$ is calculated from the relation $\vert \Gamma K\vert=\frac{4\pi}{3a}=\frac{4\pi}{\lambda_i}\sin{\theta}$.\ To map the $\Gamma M$ direction, we rotate the sample 30 degrees to align the $\Gamma M$ path with the horizontal direction as indicated in Fig. \ref{2}(f) and, from that position, we rotate the angle $\theta$.\ Here, the maximum angle is indicated by $\vert \Gamma M\vert=\frac{2\pi}{\sqrt{3}a}=\frac{4\pi}{\lambda_i}\sin{\theta}$.\ Mapping the $KM$ path requires the simultaneous variation of two specific angles $\alpha$ and $\theta$ to measure the intersecting point of the blue segment and the horizontal direction, as depicted in Fig. \ref{2}(e).\ The calculated and measured frequencies are in good agreement and we attribute the residual frequency mismatch to fabrication fluctuations and the non-vertical sidewalls.\ Some modes in Fig. \ref{2}(c) are undetectable in the experiment as their displacement is predominantly in-plane and therefore do not scatter enough light to be detected.\ The light-blue region in Fig. \ref{2}(c) highlights the mechanical gap of this particular crystal.\ Within this frequency window, no mechanical modes were measured for any angle of incidence in any high-symmetry direction, covering a broad spectral range of 5.3 GHz centered at 8.4 GHz, which corresponds to a gap to mid-gap ratio of 64\%. This crystal has the largest measured mechanical gap in the hypersonic regime to date.\ We also explore the possibility to spectrally tune the gap as a function of the geometry by variation of the lattice constant $a$.\ The band gap evolution calculated from FEM and measured spectra for crystals with periods of 220 nm, 330 nm and 440 nm, can be found in section S5 of SI.\ We confirm the spectral tunability of the gap from 4 to 12 GHz. Subsequently, tuning of phononic guided modes is also possible.

Finally, we demonstrate the possibility to create phononic waveguides with the phononic insulator presented here.\ For this, we design and fabricate a waveguide surrounded on both sides by shamrock phononic crystals with inverted symmetry as shown in the SEM image in Fig. \ref{3}(a).\ This structure has a periodicity of $a$ = 440 nm, a waveguide width of $w$ = 184 nm, and the same fill fraction and thickness as the previous structures ($r/a=0.22$, $t=220$ nm).\ The mirror symmetry of the crystal with respect to the defect line is crucial for proper band engineering of the guided mechanical modes.\ The two panels in Fig. \ref{3}(b) show the Brillouin spectra measured on the waveguide (top), and on the surrounding phononic crystal (bottom), as specified in the insets, taken at the same incident angle of $23.8^\circ$.\ The blue region indicates the phononic band gap of the structure.\ The two peaks measured within this gap in the top panel of Fig. \ref{3}(b) at 5.7 GHz and 7.1 GHz are clear experimental evidence of mechanical vibrations localized in the phononic waveguide.\ In order to detect these localized modes, it is necessary to focus the light on the waveguide with a long-working distance microscope objective to reduce the spot size of the incident light down to 1.2 $\micro$m.\ In doing so, we reduce the contribution of the Brillouin scattered signal from the crystal while increasing the contribution from the waveguide.\ For the measurements of these waveguide structures, the background is higher due to a greater collection of reflected and linearly scattered light relative to that of the 3 cm lens used in Fig. \ref{2}.\ Figure \ref{3}(c) plots the dispersion relation of the waveguide accounting for the $4^\circ$ correction of the vertical sidewalls.\ As in Fig. \ref{2}c, the color intensity of the bands corresponds to the normalized coupling coefficient for the MB perturbation.\ The fully shaded regions above and below the gap correspond to the bulk crystal modes and define the band gap edges of the structure.\ The calculated dispersion relation exhibits nine guided modes but only two (indicated with associated mode profiles) have sufficient mechanical out-of-plane displacement to potentially be detected in experiment.\ The horizontal dashed lines highlight the edges of the band gap measured in Fig. \ref{3}(b) and the black dots correspond to the frequencies of the guided modes measured at different angles.\ The angle $\theta = 23.8^\circ$ corresponds to a normalized wavevector of 1.34 (replacing $q_{\parallel} = n\pi/a$ in Eq. \ref{pmc} and solving n), or $k_x/(\pi/a) = 0.66$ over the first periodic zone of the waveguide.\ The two black dots (1 and 2) coincide with the measured frequencies and wavevector in Fig. \ref{3}(b).\ We assume that measured peaks around 7 GHz corresponds with the darker flat mode around 7.7 GHz and not with the lighter curve that it is spectrally closer to.\ There is a difference of approximately 700 MHz for this band while the other peaks agree very closely with the calculated band.\ The detection of these two modes is a clear fingerprint of the existence of localized modes along the Shamrock waveguide.

In summary, we provide direct experimental evidence of the complete absence of mechanical vibrations at room temperature within a full phononic band gap that is 5.3 GHz wide with a central frequency of 8.4 GHz, measured using Brillouin light scattering.\ This measured mechanical gap is significantly wider than previous demonstrations in literature, especially at this frequency regime.\ Our approach incorporates a geometrical pattern that distributes the mass within the unit cell, forming mass clusters connected by narrow necks that result in the destructive interference of phonon waves, giving rise to these wide mechanical gaps.\ We achieve control over the width and frequency of the gap by fine-tuning of the geometrical parameters of the structure, enabling spectral tunability of the gap from 4 GHz to 12 GHz.\ This tunability is extended to the guided modes of a line-defect waveguide, enabling engineering of the frequency and number of localized modes within the structure.\ This is the first demonstration of mechanical guided modes at hypersonic frequencies in the GHz regime using line-defect waveguides, measured at room temperature without any external excitation.\ The structures are fabricated on the standard SOI platform which enables facile integration into existing photonic systems.\ The hypersonic insulator presented here also is a photonic insulator for Transverse-Electric (TE) modes at telecom wavelengths \cite{ immo, arreguiprb} and can be used to simultaneously engineer phononic and photonic transport enhancing the optomechanical coupling between THz photons and GHz phonons.\ This makes the crystal an ideal transducer in photonic circuits with potential applications in high-speed signal processing \cite{signalprocess}.\ Furthermore, this platform can be used in applications and physical processes in which a wide mechanical band gap is required to isolate the system from thermal damping, such as in quantum cavity optomechanics or organic molecular systems \cite{emitters}.

%--------------------------------------------------------------------

\begin{center}
\textbf{Acknowledgements}
\end{center}

This project has received funding from the European Union's H2020 FET Proactive project TOCHA (No. 824140) and Horizon 2020 research and innovation programme under the Marie Sklodowska-Curie grant agreement No. 754558. The ICN2 authors acknowledge funding by the Severo Ochoa program from Spanish MINECO (No. SEV-2019-0706), Plan Nacional RTI2018-093921-A-C44 (SMOOTH), and MCIN project SIP (PGC2018-101743-B-100), as well as by the CERCA Programme Generalitat de Catalunya.\ O.F. and G.A. are supported by BIST PhD. Fellowships, R.C.N. by a Marie Sklodowska-Curie fellowship (No. 897148), and P.D.G. by a Ramon y Cajal fellowship No. RyC-2015-18124. M.A. and S.S. gratefully acknowledge funding from the Villum Foundation Young Investigator Program (No.\ 13170), the Danish National Research Foundation (No.\ DNRF147 - NanoPhoton), Innovation Fund Denmark (No.\ 0175-00022 - NEXUS), and Independent Research Fund Denmark (Grant No.\ 0135-00315 - VAFL).

\begin{center}
\textbf{Author contributions}
\end{center}
 O.F. designed, simulated, and characterized the samples. M.A. and S.S. fabricated the samples. G.A., R.C.N., and J.G.B. contributed to the data analysis. C.M.S.T. and P.D.G. supervised the work. P.D.G. conceived the idea and the project. O.F. and P.D.G. wrote the manuscript with contributions and input from all authors.

\begin{center}
\textbf{Competing financial interests}
\end{center}
The authors declare no competing financial interests.

\begin{center}
\textbf{Methods}
\end{center}
Methods are available in the supplementary information.

\begin{center}
\textbf{Data availability}
\end{center}
Data is available upon reasonable request.

%------------------------------- sup info -------------------
%\large

\begin{widetext}

\newpage

\renewcommand{\figurename}{\textbf{Figure}}
\makeatletter
\renewcommand{\thefigure}{S\@arabic\c@figure}
\makeatother
\renewcommand\theequation{S\arabic{equation}}
\renewcommand\thetable{S\arabic{table}}
\renewcommand{\bibname}{References}

\section*{SUPPLEMENTARY INFORMATION}

\setcounter{figure}{0}
\setcounter{equation}{0}

\section*{S1. CRYSTAL DESIGN, FABRICATION AND CHARACTERIZATION}
\label{fabrication}

\subsection*{Shamrock design}

The periodic pattern of the phononic crystal consists of three tangential circles of radius $r= 0.22a$, where $a$ is the period of the structure, shifted by a distance $f= 2r/\sqrt{3}$ (Fig. \ref{SI_shamrock}) from the common center and arranged to form what we call a shamrock~\cite{Soellner}. Three small semicircles of radius $f-r$ have been added in the junctions to smooth out any sharp corners in the geometry and create a structure that can be fabricated.

\begin{figure}[h!]
 \centering
  \includegraphics[width=0.8\columnwidth]{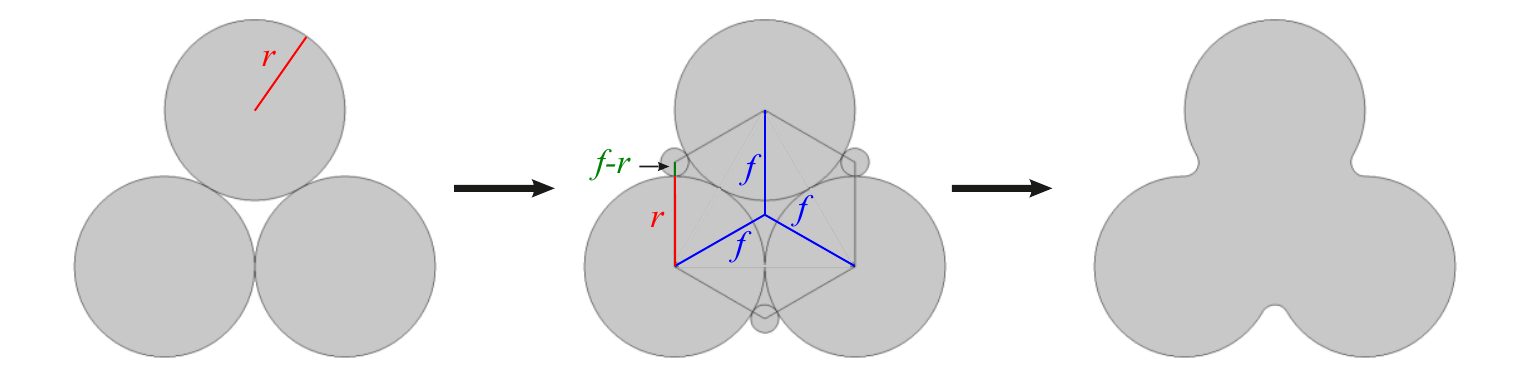}
\caption{\textbf{Shamrock geometry} Three adjacent circles of radius $r$ form the replication pattern of the shamrock phononic crystal.}
\label{SI_shamrock}
\end{figure}

\subsection*{Fabrication}

All the structures studied here were fabricated in chips cleaved from a commercial 12-inch silicon-on-insulator (SOI) wafer, which has a nominal 220 nm thick device layer and a 2 $\micro$m buried oxide. The lithography, dry-etching, and release etch of the suspended membrane is carried out as detailed in \cite{albrechtsen2021} with some minor changes. Since we do not need to make as small features, we spin-coat a thicker (180 nm) softmask (chemically semi-amplified resist, CSAR6200.09), which enables us to reduce the periodic sidewall roughness (scallops). We define the patterns with a 100 keV 100 MHz JEOL-9500FSZ electron-beam writer with a current 200 pA and a dose density of 2.5 aC/nm$^2$ in the center, boosted by 10 \% in the corners to compensate long-range proximity effects across the large crystals \cite{owen_proximity_1983}. We consider the variance of the long-range effects, $\beta$ = 30$\micro$m, and the relative weight compared to the forward scattering, $\eta=0.5$. We further modify the cyclic dry-etching process CORE \cite{nguyen2020}, with 3 changes from the process in Ref. \cite{albrechtsen2021}. Specifically, we increase the number of cycles to 14 and reduce the time of the E-step (etch) to 45 s. to reduce the size of the scallops. In order to reduce sidewall erosion, we increase the time of the O-step (passivation) to 5 s. The structures are released by a 2.8 $\micro$m isotropic underetch of the buried oxide \cite{albrechtsen2021}.

Figure \ref{SI_fab} shows the fabricated samples with periods $a$ of 220, 330 and 440 nanometers with the same radius $r=0.22a$. All the crystals have a full area of 50x50 $\micro m$.

\begin{figure}[h!]
 \centering
  \includegraphics[width=\columnwidth]{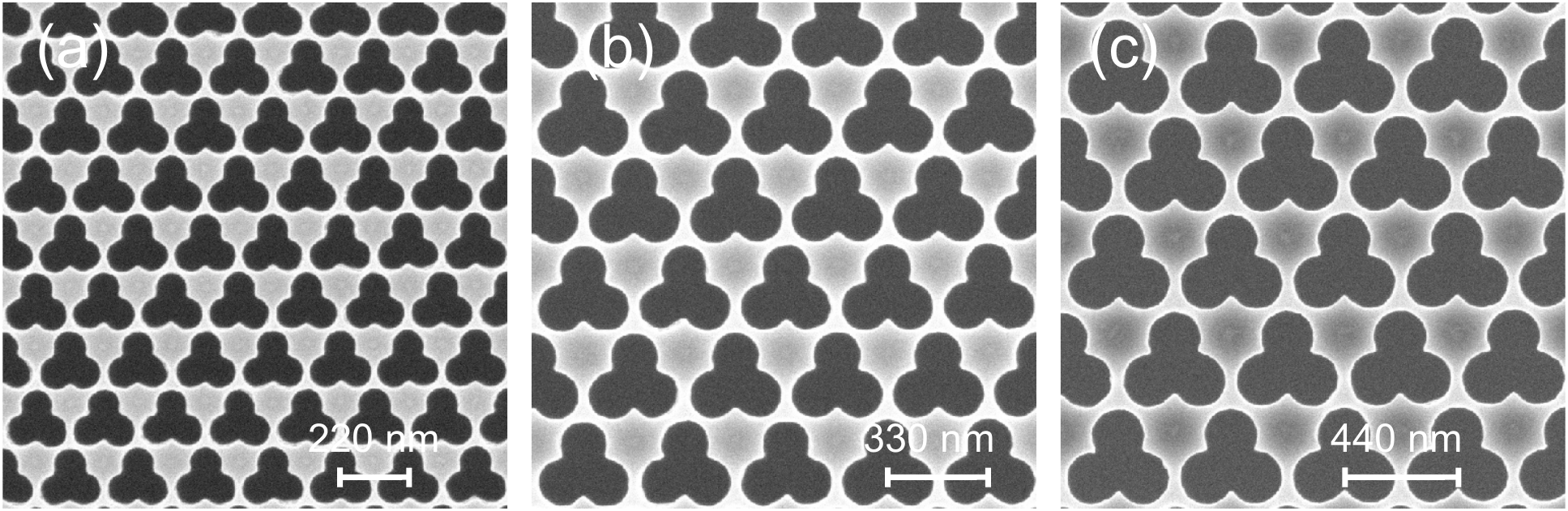}
\caption{\textbf{Fabricated samples} Scanning-electron micrographs of the shamrock phononic crystals with period $a$ of (a) 220 nm, (b) 330 nm, and (c) 440 nm.}
\label{SI_fab}
\end{figure}

\subsection*{Contour fitting and statistical analysis}

The average shape of the fabricated shamrock holes is obtained from an ensemble of 56 holes extracted from the same SEM image.\ Each contour comes from a binarized version of the original SEM image adjusted to obtain the most accurate result as shown Fig. \ref{SI_cont}(a).\ Figure \ref{SI_cont}(b) plots the ensemble of all contours (red dots) and its mean value (blue dots). The shape of the individual fitted shamrocks are very similar which is a qualitative measure of the fabrication tolerance. To quantify the degree of imperfection, we compare the mean value and standard deviation of the ensemble averaged area of the shamrocks with the nominal value. Fig. \ref{SI_cont}(c) plots the histogram distribution of the areas in nm$^2$ of each individual shamrock (in red), and the mean value (in blue). The gray line in Fig. \ref{SI_cont}(c) indicates the nominal area of the designed geometry. The difference between the average area, $5.12\times10^4 \text{nm}^2$, and the nominal area, $4.91\times10^4 \text{nm}^2$, is attributed to the not fully resolved corners of the shamrock, as displayed in the inset of the figure. This clearly shows the significance of avoiding sharp corners in the unit cell geometry, which requires a much accurate fabrication process. Although geometric inspection using SEM images has its limitations, this simple analysis shows the high quality of the fabrication process used here and also the simplicity and robustness of the pattern selected to create the periodicity.

\begin{figure}[h!]
 \centering
  \includegraphics[width=\columnwidth]{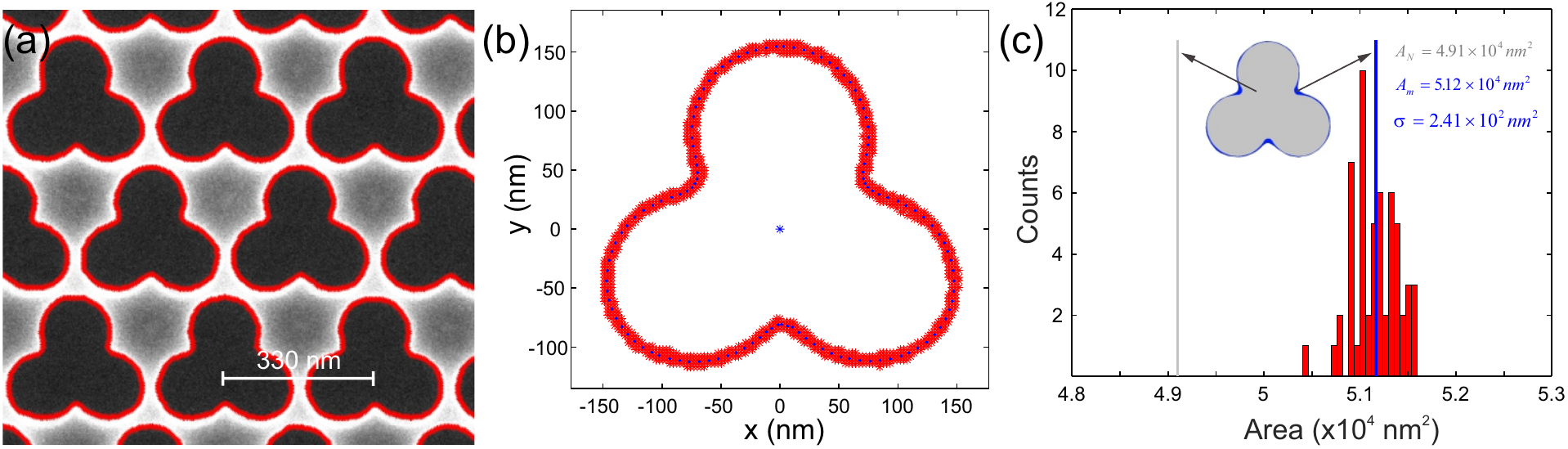}
\caption{\textbf{Fitting and statistical analysis} (a) Fitted contours in a shamrock crystal with a period of 330 nm. (b) Superposition of all the contours (red) and its mean value (blue points). (c) Histogram for the fitted areas and its mean, compared with the nominal value of the designed area. There is a small difference caused by the etching in the intersection of the circles.}
\label{SI_cont}
\end{figure}

\newpage

\section*{S2. Band structure of the irreducible Brillouin zone}
\label{bands}

A simple periodic geometry can be created in a material using electron-beam lithography by etching circularly symmetric holes (C1), often in a hexagonal lattice (C6v) \cite{Sakoda}, leading to crystals that belong to the planar space-group (also called wallpaper group) p6mm. However, different (more restricted) symmetries of the pattern in the unit cell and the symmetries of the material at the atomic scale might induce a discrete set of rotational symmetries in addition to their translation invariance. Figure \ref{SI_irreducible}(a-c) illustrates how the existing symmetries of the unit cell of a triangular lattice give rise to reduced sizes of the Brillouin zone (BZ) via some examples, where the top panel represents the real-space Wigner-Seitz cell and the bottom panel the corresponding first BZ. The circle in Fig. \ref{SI_irreducible}(a) possesses full rotational symmetry (C1) and the 12 operations that map the hexagonal Wigner-Seitz cell to itself are preserved, leading to an irreducible Brillouin zone (IBZ) (the shaded region inside $\Gamma KM$) of size 1/12, in units of the first BZ. The equilateral triangle in \ref{SI_irreducible}(b) breaks the inversion symmetry (C2) and three mirror symmetries ($\widehat{\sigma}_{y}$ , $\widehat{\sigma}_{y'}$ and $\widehat{\sigma}_{y''}$) leading to an IBZ two times larger.\ Other geometries such as a square can preserve inversion symmetry but break a higher number of mirror symmetries, leading to an IBZ one quarter of the first BZ in size (c). The shamrock pattern used here \cite{Soellner}, preserves the symmetry described in \ref{SI_irreducible}(b). However, if we consider the anisotropy of the silicon stiffness tensor, the mirror symmetry of the unit cell is broken resulting in the symmetry described in \ref{SI_irreducible}(c). This leads to a IBZ given by the region bounded by the path $\Gamma KMK_1M_1\Gamma$. Figure \ref{SI_irreducible}(d) plots the phononic dispersion relation of the system presented in Fig. 1 of the main text along this IBZ. For simplicity, in our experiment we measure along the simplified path $\Gamma KM\Gamma$ indicated in Fig. \ref{SI_irreducible}(b). This is sufficient to characterize the full width of the phononic gap as its limiting frequencies do not change with respect to the IBZ.
\begin{figure}[h!]
 \centering
  \includegraphics[width=\columnwidth]{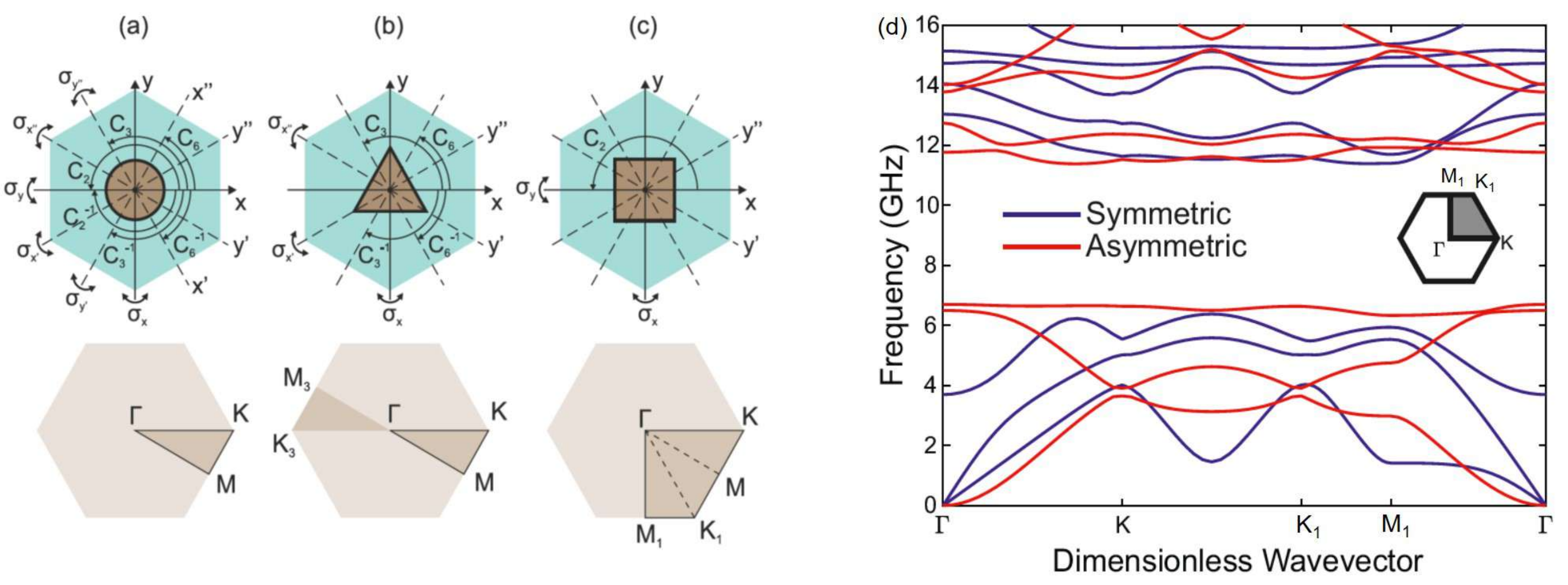}
\caption{\textbf{Irreducible Brillouin Zone (IBZ)}. The symmetries of the unit cell pattern that map the lattice into itself are called the point group of the crystal and define the IBZ, the minimum region in k-space that needs to be sampled to access the full eigenspectrum $\omega_n(k)$. (a) A circle preserves all mirror symmetries $\sigma_{xx';x'';yy';y''}$, all rotational symmetries $C_{\pm 6; \pm 3;\pm 2}$ and the inversion symmetry $C_{\pm 2}$ of the hexagonal Wigner-Seitz (WS) cell of a triangular lattice, leading to a IBZ delimited by $\Gamma KM$ IBZ, shaded in brown. (b) An equilateral triangle breaks the $\sigma_{x';y;y'}$ mirror symmetries and the inversion symmetry $C_{\pm 2}$, which leads to the shaded region. Use of time-reversal invariance finally leads to the same $\Gamma KM$ IBZ. Larger IBZs occur when the pattern breaks additional symmetries, as does (c) a square leading to an IBZ of size 1/4 in units of the 1BZ area. (d) Calculated mechanical dispersion relation along the the borders of the irreducible Brillouin zone $\Gamma KMK_1M_1\Gamma$ for the phononic crystal presented in Fig. 1a of the main text. Blue and red curves represents symmetric and asymmetric modes with respect to the mid plane of the slab.}
\label{SI_irreducible}
\end{figure}

\newpage

\section*{S3. Brillouin light scattering spectroscopy}
\label{blss}

The setup used to measure the spontaneous backward Brillouin scattering was a tandem Fabry-Perot interferometer. A 532 nm continuous-wave solid state laser with a linewidth of 5 MHz was used for the incident and reference beam. The scattered light is routed into a six-times pass Fabry-Perot interferometer that is comprised of two cavities in series, FP1 and FP2 in Fig. \ref{SI_setup}. This system increases the spectral resolution and allows high stability and accurate measurements of spectra over long acquisition times. For measurements on the crystal, the incident power was 2.5 mW using a 3 cm focal lens, with an acquisition time of 24 hours for each measured angle. For measurements on the waveguide and the surrounding shield, the incident power was 1 mW using a long working distance objective lens with 100x magnification, also with an acquisition time of 24 hours for each spectrum.
\begin{figure}[h!]
 \centering
  \includegraphics[width=0.8\columnwidth]{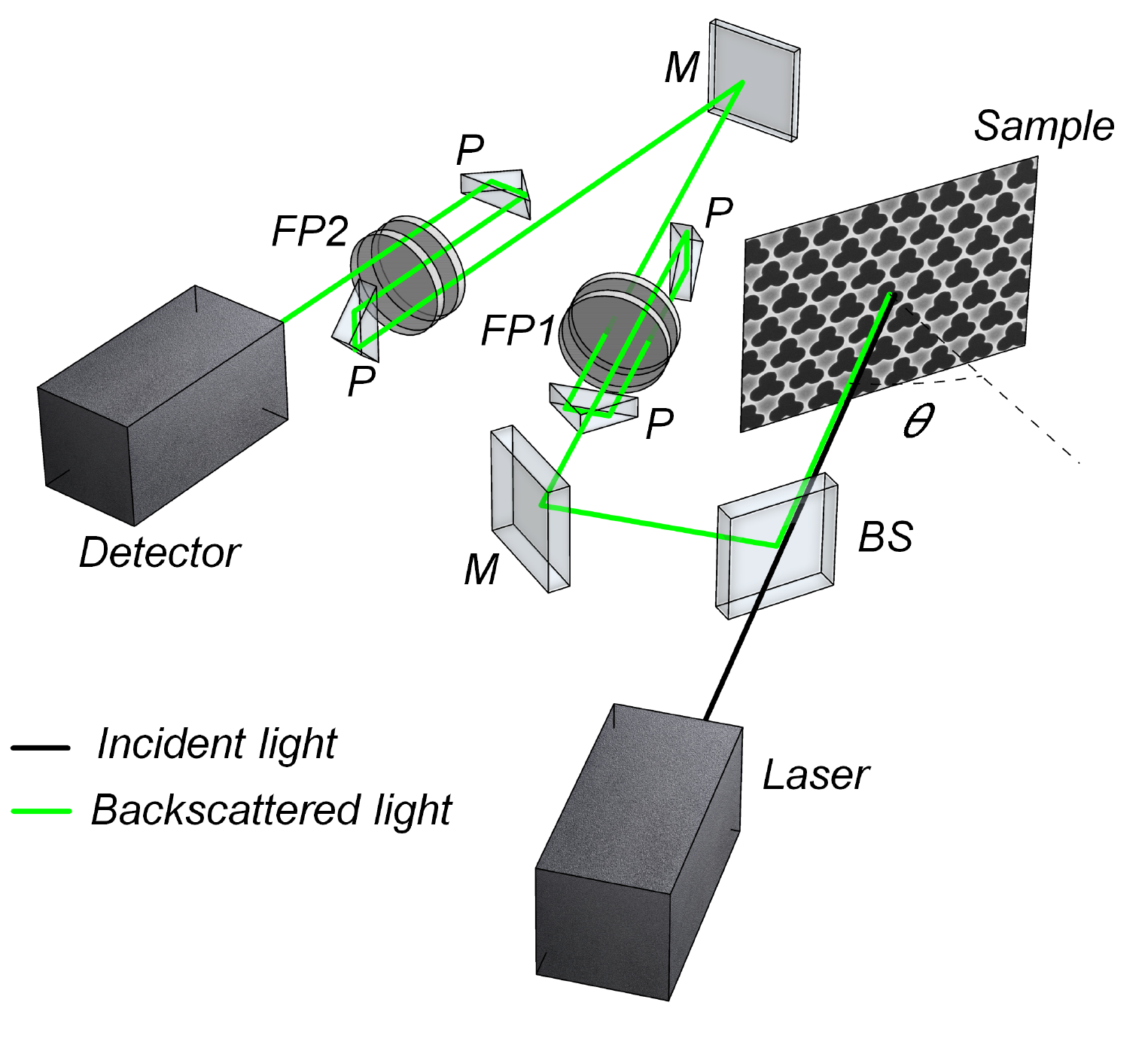}
\caption{\textbf{Brillouin light scattering spectroscopy}. simplified scheme of the tandem Fabry-Perot interferometer used to measure and enhance the backscattered signal. BS: beam spliter, M: mirror, P: prism, FP: Fabry-Perot cavity.}
\label{SI_setup}
\end{figure}
\newpage
\section*{S4. SCATTERING EFFICIENCY}
\label{efficiency}
The scattering efficiency quantifies how much a mechanical mode scatters the incident light. To calculate it, we need to consider the origin of the periodic perturbation of the refractive index. It is caused by two main physical mechanisms. The most well-known is the photoelastic effect (PE) induced by the elastic strain of the material~\cite{nelson1971theory}, and the moving boundary effect (MB) induced by the mechanical displacement of the surface~\cite{johnson2002perturbation}. The latter becomes important when light is on the surface, such as in our case. Here we made a first approximation taking into account only the MB perturbation and neglecting the PE contribution, given the small volume of interaction in our experiment. This approximation allows us to obtain the coupling coefficients in an almost fully analytical way.  According to \cite{johnson2002perturbation}, the MB coupling is given by:
\begin{equation}
\label{eq:se1}
\kappa_{mb}=\int_A U_z \left[ \Delta\epsilon_{12} \left(\textbf{E}_{s,\parallel}^*\cdot\textbf{E}_{i,\parallel}\right)-\Delta\epsilon_{12}^{-1}\left(\textbf{D}_{s,\perp}^*\cdot\textbf{D}_{i,\perp}^{'}\right)\right]dA
\end{equation}
where $U_z$ is the normal displacement to the surface $A$ as indicated in Fig. \ref{SI_scatteff}(a).  $\Delta\epsilon_{12}=(\epsilon_1-\epsilon_2)$ and $\Delta\epsilon_{12}^{-1}=(\epsilon^{-1}_1-\epsilon^{-1}_2)$,  $\textbf{E}$ and $\textbf{D}$ are electric and displacement fields, and scripts $s$ and $i$ denote scattered and incident fields respectively. Taking into account the relation between the relative permittivity and the refractive index, $\epsilon=\epsilon_0n^2$, we can write
\begin{equation}
\label{eq:se2}
\Delta\epsilon_{12}=\epsilon_0(n_1^2-n_2^2)
\end{equation}
\begin{equation}
\label{eq:se3}
\Delta\epsilon_{12}^{-1}=\frac{1}{\epsilon_0}\left(\frac{1}{n_1^2}-\frac{1}{n_2^2}\right).
\end{equation}
Replacing the normal component of the displacement field by the electric field component, $\textbf{D}=\epsilon\textbf{E}$, we can rewrite Eq. \ref{eq:se1} as
\begin{equation}
\label{eq:se4}
\kappa_{mb}=\int_A U_z\epsilon_0 \left[ \left(n_1^2-n_2^2\right) \left(\textbf{E}_{s,\parallel}^*\cdot\textbf{E}_{i,\parallel}\right)-n_2^4\left(\frac{1}{n_1^2}-\frac{1}{n_2^2}\right)\left(\textbf{E}_{s,\perp}^*\cdot\textbf{E}_{i,\perp}^{'}\right)\right]dA.
\end{equation}
The incident and scattered electric field in the sample surface can be treated as planes waves propagating in free space and can be written as:
\begin{equation}
\label{eq:se5}
\textbf{E}_i(r,t)=\frac{1}{2}\textbf{E}_i(x,y)e^{-i(\omega_it-\textbf{k}_i\cdot\textbf{z})}+c.c.
\end{equation}
\begin{equation}
\label{eq:se6}
\textbf{E}_s(r,t)=\frac{1}{2}\textbf{E}_s(x,y)e^{-i(\omega_st+\textbf{k}_s\cdot\textbf{z})}+c.c.
\end{equation}
We need to account for the polarization of the electric field used in the experiment (TM polarization) as depicted in Fig. \ref{SI_scatteff}. For the backward Brillouin scattering configuration, $\vec{k}_s=-\vec{k}_i$ and $\omega_s=\omega_i\pm\Omega$ or $\omega_s\approx\omega_i$ given the small frequency of mechanical modes (GHz) compared with the incident light frequency (THz). $\textbf{E}_s(x,y)=\textbf{E}_i(x,y)=\textbf{E}_i$ because light is propagating in free space. Therefore the product of the incident and scattered fields can be simplified to $\textbf{E}^*_s\cdot\textbf{E}_i=E^2_i$. Incident and scattered light can then be treated as the same fields.
\begin{figure}[h!]
 \centering
  \includegraphics[width=1\columnwidth]{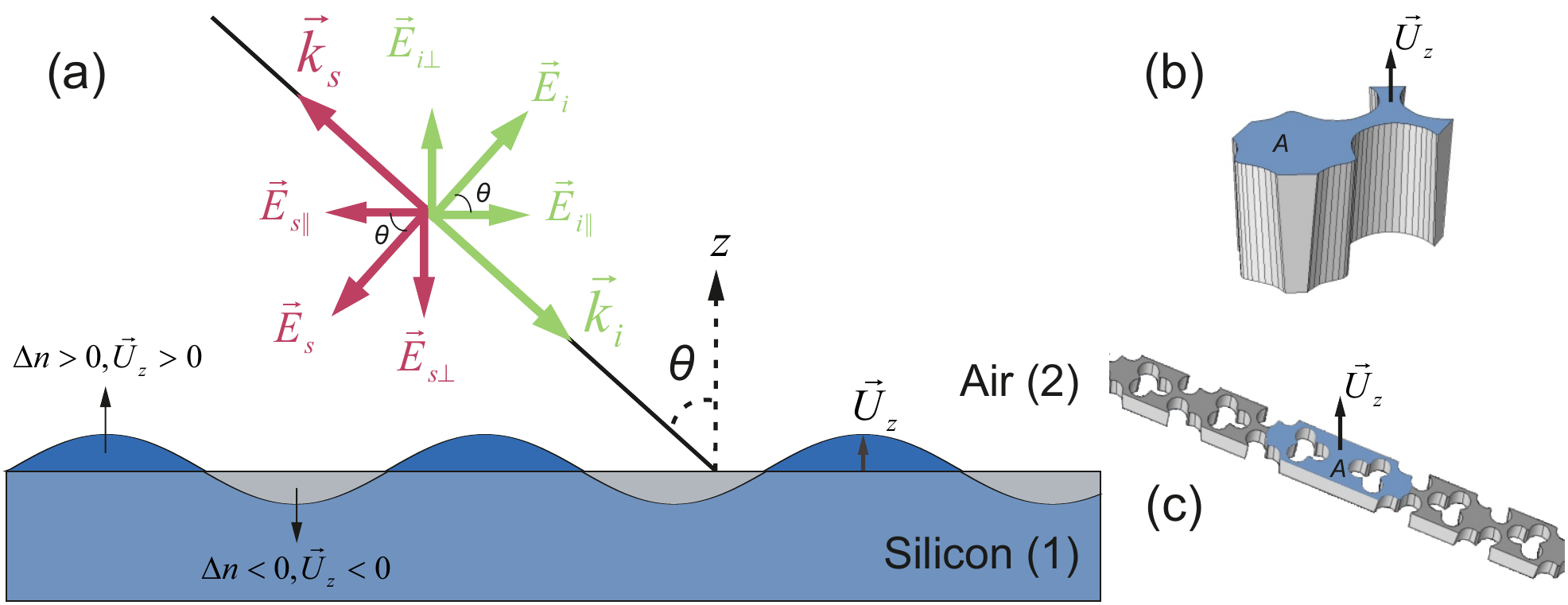}
\caption{\textbf{Optomechanical coupling mediated by the moving-boundary effect.} (a) Components of the incident and backscattered electromagnetic field (green and red respectively). Each mode of the structure leads to a particular vertical displacement of the surface that determines the scattering contribution. (b) and (c) highlight the surface to be integrated in the phononic crystal and phononic waveguide respectively.}
\label{SI_scatteff}
\end{figure}
As can be seen in Fig. \ref{SI_scatteff}(a), we have $\textbf{E}_{s,\parallel}=-\textbf{E}_{i,\parallel}$ and $\textbf{E}_{s,\perp}=-\textbf{E}_{i,\perp}$, where the parallel and perpendicular components of electric field are given by $\textbf{E}_{i,\parallel}=\textbf{E}_i\cos{\theta}$, and $\textbf{E}_{i,\perp}=\textbf{E}_i\sin{\theta}$. Replacing all these expressions in Eq.~\ref{eq:se4}, we can write
\begin{equation}
\label{eq:se7}
\kappa_{mb}=\int_A U_z\epsilon_0 \left[ \left(n_1^2-n_2^2\right) \left(-E_i^2\cos^2{\theta}\right)-\left(n_2^2-n_1^2\right)\frac{n_2^2}{n_1^2}\left(-E_i^2\sin^2{\theta}\right)\right]dA.
\end{equation}
The entire expression inside brackets can be taken out the integral. We only need to account for the change of the refractive index with respect to the surface displacement. To do so, we always assume a positive change of the refractive index and take the absolute value of the normal displacement $\textbf{U}_z$.
\begin{equation}
\label{eq:se8}
\kappa_{mb}=-\epsilon_0 E^2_i\left(n^2_1-n^2_2\right)\left[\cos^2{\theta}+\frac{n^2_2}{n^2_1}\sin^2{\theta}\right]\int_A \vert U_z\vert dA.
\end{equation}
Here the coupling coefficient does not depend on the incident field and it can be normalized to $E^2_i=1W$. Each mechanical mode has its own scattering efficiency depending on the surface displacement $U_z$. The bigger the refractive index contrast, the bigger the MB contribution. Finally, the scattering cross section needs to be accounted for. The illuminated area on the sample surface changes with the angle, being a circle of area $A_c=\pi b^2$ for $\theta=0$, where b is the beam diameter, and an ellipse with axes $b$ and $l=b/cos(\theta)$, and area $A_e=\pi b l$ for a certain angle $\theta$. Therefore, Eq.~\ref{eq:se8} needs to be divided by $cos(\theta)$ to account for the scattering cross section. Finally, replacing $n_1=n_{si}$ and $n_2=1$, the coupling coefficient for the MB perturbation is given by
\begin{equation}
\label{eq:se9}
\kappa_{mb}=-\epsilon_0 E^2_i\left(n_{Si}^2-1\right)\left[\cos{\theta}+\frac{1}{n_{Si}^2}\tan{\theta}\sin{\theta}\right]\int_A \vert U_z\vert dA.
\end{equation}
For the phononic crystal, the area integral has to be computed over the top face of the unit cell as indicated by Fig. \ref{SI_scatteff}(b). For the waveguide, we are interested in the localized modes; therefore the area integral is not calculated in the unit entire cell b  ut only in the central region as indicated in Fig. \ref{SI_scatteff}(c). Although the expression presented here does not account for the PE contribution, it is a good approximation to determine if a particular mechanical mode can be detected in our experimental setup. Equation \ref{eq:se9} is used to calculate the color intensity of bands in Fig. 2(c) and Fig. 3(c) in the main text.
\newpage
\section*{S5. SPECTRAL TUNABILITY}
\label{tunability}

We explored the geometrical tunability of the gap induced by the variation of the lattice constant $a$ (see inset of Fig.~\ref{gapmap}(a)).\ The spectral evolution of the gap width as a function of period $a$ in Fig. \ref{gapmap}(a) was calculated with a finite element method assuming a slab thickness of 220 nm and a unit cell fill fraction of $r/a$ = 0.22.\ Here we do not account for or apply any correction to the vertical profile of the crystal.\ The gap opens for a lattice unit of $a$ = 100 nm and has a maximum of almost 6 GHz for $a$ = 300 nm.\ We confirm the mechanical spectral shift by measuring the Brillouin inelastic scattering spectrum at the high-symmetry point $K$ for three different structures fabricated on silicon-on-insulator with different periods of $a$ = 220 nm, 330 nm and 440 nm, as shown Fig. \ref{gapmap}(b).\ We measure at angles which correspond to $K$ at $\theta = 53.7^\circ$, $32.5^\circ$, and $23.8^\circ$, respectively.\ The agreement between simulations and experiment in Fig. \ref{gapmap} is strong evidence of the gap displacement.\ As mentioned previously, differences in frequency between theory and experiment are due to unavoidable fluctuations in the fabrication process which become more significant for smaller lattice units.\ The broad tunability of the mechanical gaps measured here, from 4 GHz to 12 GHz, enables a fine control of hypersonic phonon routing and cavity optomechanics based on the shamrock geometry.
\begin{figure}[h!]
\includegraphics[width=0.7\columnwidth]{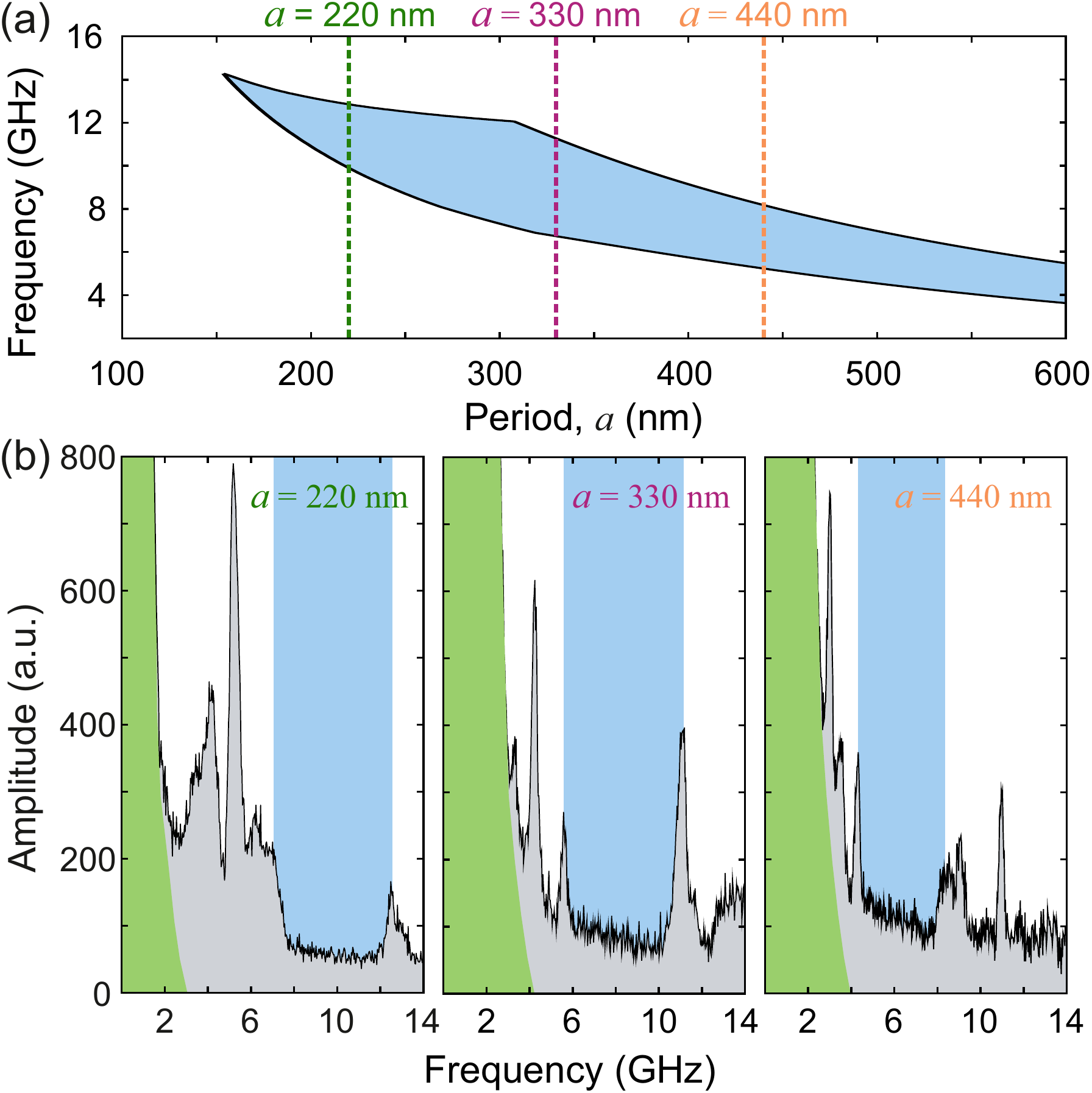}
\caption{ \label{gapmap} \textbf{Geometric tunability of the system} (a) Spectral evolution of the mechanical gap as a function of the lattice period $a$, considering the same unit cell fill fraction, $r/a = 0.22$. (b) Brillouin light scattering spectra measured for three different crystals with periods $a$ = 220 nm, 330 nm, and 440 nm. The light-blue regions highlight the spectral position of the mechanical gap in each structure.}
\end{figure}
\end{widetext}


\begin{thebibliography}{99}
%1
\bibitem{accelerometer}
A. G. Krause, M. Winger, T. D. Blasius, Q. Lin and O. Painter. A high-resolution microchip optomechanical accelerometer. Nat. Photon. \textbf{6}, 768 (2012).
%2
\bibitem{masssensor}
J. Chaste et al. A nanomechanical mass sensor with yoctogram resolution. Nat. Nanotechnol. \textbf{7}, 301 (2012).
%3
\bibitem{forcesensor}
E. Gavartin, P. Verlot and T. J. Kippenberg. A hybrid on-chip optomechanical transducer for ultrasensitive force measurements. Nat. Nanotechnol. \textbf{7}, 509 (2012).
%4
\bibitem{sidebandcool}
J. D. Teufel et al. Sideband cooling of micromechanical motion to the quantum ground state. Nature \textbf{475}, 359 (2011).
%5
\bibitem{lasercool}
J. Chan et al. Laser cooling of a nanomechanical oscillator into its quantum ground state. Nature \textbf{478}, 89 (2011).
%6
\bibitem{sigalas}
M. Sigalas and E. N. Economou. Band structure of elastic waves in two dimensional systems. Solid State Commun. \textbf{86}, 141 (1993).
%7
\bibitem{kushawa}
M. S. Kushwaha, P. Halevi, L. Dobrzynski, and B. Djafari-Rouhani. Acoustic band structure of periodic elastic composites. Phys. Rev. Lett. \textbf{71}, 2022 (1993).
%8
\bibitem{martinez}
R. Mart\'inez-Sala et al. Sound attenuation by sculpture. Nature \textbf{378}, 241 (1995).
%9
\bibitem{gorishnyy}
T. Gorishnyy, C. K. Ullal, M. Maldovan, G. Fytas, and E. L. Thomas. Hypersonic Phononic Crystals. Phys. Rev. Lett. \textbf{94}, 115501 (2005).
%10
\bibitem{zen}
N. Zen, T. A. Puurtinen, T. J. Isotalo, S. Chaudhuri and I. J. Maasilta. Engineering thermal conductance using a two-dimensional phononic crystal. Nat. Commun. \textbf{5}, 3435 (2014).
%11
\bibitem{eichenfield}
M. Eichenfield, J. Chan, R. M. Camacho, K. J. Vahala and O. Painter. Optomechanical crystals. Nature \textbf{462}, 78 (2009).
%12
\bibitem{rouhani}
B. Djafari-Rouhani, S. El-Jallal and Y. Pennec. Phoxonic crystals and cavity optomechanics. C. R. Phys. \textbf{17}, 555 (2016).
%13
\bibitem{maccabe}
G. S. MacCabe et al. Nano-acoustic resonator with ultralong phonon lifetime. Science \textbf{370}, 840 (2020).
%14
\bibitem{kfang}
K. Fang, M. H. Matheny, X. Luan and O. Painter. Optical transduction and routing of microwave phonons in cavity-optomechanical circuits. Nat. Photon. \textbf{10}, 489 (2016).
%15
\bibitem{rnpatel}
R. N. Patel et al. Single mode phononic wire. Phys. Rev. Lett. \textbf{121}, 040501 (2018).
%
%\bibitem{snowflake1}
%A. H. Safavi-Naeini et al. Two-Dimensional Phononic-Photonic Band Gap Optomechanical Crystal Cavity. Phys. Rev. Lett. \textbf{112}, 153603 (2014).
%16
\bibitem{snowflake2}
H. Ren et al. Two-dimensional optomechanical crystal cavity with high quantum cooperativity. Nat. Commun. \textbf{11}, 3373 (2020).
%17
\bibitem{jordinc}
J. Gomis-Bresco et al. A one-dimensional optomechanical crystal with a complete phononic band gap. Nat. Commun. \textbf{5}, 4452 (2014).
%18
\bibitem{mohammadi}
S. Mohammadi, A. A. Eftekhar, A. Khelif, W. D. Hunt and Ali Adibi. Evidence of large high frequency complete phononic band gaps in silicon phononic crystal plates. Appl. Phys. Lett. \textbf{92}, 221905 (2008).
%19
\bibitem{soliman}
Y. M. Soliman et al. Phononic crystals operating in the gigahertz range with extremely wide band gaps. Appl. Phys. Lett. \textbf{97}, 193502 (2010).
%20
\bibitem{gorisse}
M. Gorisse et al. Observation of band gaps in the gigahertz range and deaf bands in a hypersonic aluminum nitride phononic crystal slab. Appl. Phys. Lett. \textbf{98}, 234103 (2011).
%21
\bibitem{cheng}
W. Cheng, J. Wang, U. Jonas, G. Fytas and N. Stefanou. Observation and tuning of hypersonic bandgaps in colloidal crystals. Nat. Mater. \textbf{5}, 830 (2006).
%22
\bibitem{bart1}
B. Graczykowski et al. Phonon dispersion in hypersonic two-dimensional phononic crystal membranes. Phys. Rev. B \textbf{91}, 075414 (2015).
%23
\bibitem{moli}
Q. Liu, H. Li and M. Li. Electromechanical Brillouin scattering in integrated optomechanical waveguides. Optica \textbf{6}, 778 (2019).
%24
\bibitem{immo}
I. S\"ollner, L. Midolo, and P. Lodahl. Deterministic Single-Phonon Source Triggered by a Single Photon. Phys. Rev. Lett. \textbf{116}, 234301 (2016).
%25
\bibitem{arreguiprb}
G. Arregui, D. Navarro-Urrios, N. Kehagias, C. M. Sotomayor Torres, and P. D. Garc\'ia. All-optical radio-frequency modulation of Anderson-localized modes. Phys. Rev. B \textbf{98}, 180202(R) (2018).
%26
\bibitem{comsol}
COSMOL Inc. COMSOL Multiphysics 5.1 http://www.comsol.com/ (2022).
%27
\bibitem{snowflake0}
A. H. Safavi-Naeini and O. Painter. Design of optomechanical cavities and waveguides on a simultaneous bandgap phononic-photonic crystal slab. Opt. Express \textbf{18}, 14926 (2010).
%28
\bibitem{abalandin}
F. Kargar and A. A. Balandin. Advances in Brillouin-Mandelstam light-scattering spectroscopy. Nat. Photon. \textbf{15}, 720 (2021).
%29
\bibitem{carlotti}
G. Carlotti. Elastic Characterization of Transparent and Opaque Films, Multilayers and Acoustic Resonators by Surface Brillouin Scattering: A Review. Appl. Sciences \textbf{8}, 124 (2018).
%30
\bibitem{boyd}
R. W. Boyd, Nonlinear Optics 3rd edn. (Academic Press, 2008).
%31
\bibitem{johnson}
S. G. Johnson et al. Perturbation theory for Maxwell's equations with shifting material boundaries. Phys. Rev E. \textbf{65}, 066611 (2002).
%32
\bibitem{vanlaer}
R. Van Laer, B. Kuyken, D. Van Thourhout and R. Baets. Interaction between light and highly confined hypersound in a silicon photonic nanowire. Nat. Photon. \textbf{9}, 199 (2015).
%33
\bibitem{selfcancel}
O. Florez et al. Brillouin scattering self-cancellation. Nat. Commun. \textbf{7}, 11759 (2016).
%34
\bibitem{jcuffe}
J. Cuffe et al. Phonons in Slow Motion: Dispersion Relations in Ultrathin Si Membranes. Nano Letters \textbf{12}, 3569 (2012).
%35
\bibitem{brillouin}
L. Brillouin. Diffusion de la lumi{\`e}re et des rayons X par un corps transparent homog{\`e}ne. Ann. Phys. \textbf{9}, 17. (1922).
%36
\bibitem{loudon}
R. Loudon and J. R. Sandercock. Analysis of the light-scattering cross section for surface ripples on solids. J. Phys. C: solid State Phys. \textbf{13}, 2609 (1980).
%37
\bibitem{signalprocess}
H. Shin et al. Control of coherent information via on-chip photonic-phononic emitter-receivers. Nat. Commun. \textbf{6}, 6427 (2015).
%38
\bibitem{emitters}
B. Gurlek, V. Sandoghdar, D. Martin-Cano. Engineering long-lived vibrational states for an organic molecule. Phys. Rev. Lett. \textbf{127}, 123603 (2021).

\scriptsize

\end{thebibliography}

\begin{thebibliography}{11}

\bibitem{Soellner}
I. S\"ollner, L. Midolo, and P. Lodahl. Deterministic Single-Phonon Source Triggered by a Single Photon. Phys. Rev. Lett. \textbf{116}, 234301 (2016).

\bibitem{albrechtsen2021}
M. Albrechtsen et al. Nanometer-scale photon confinement inside dielectrics. See preprint at arXiv:2108.01681 (2021).

\bibitem{owen_proximity_1983}
G. Owen and P. Rissman. Proximity effect correction for electron beam lithography by equalization of background dose. J. Appl. Phys. \textbf{54}, 3573 (1983).

\bibitem{nguyen2020}
V. T. H. Nguyen et al. The CORE sequence: a nanoscale fluorocarbon-free silicon plasma etch process based on SF6/O2 cycles with excellent 3D profile control at room temperature. ECS J. Solid State Sci. Technol. \textbf{9}, 024002 (2020).

\bibitem{Sakoda}
K. Sakoda, Optical Properties of Photonic Crystals, 2nd ed., edited by W. Rhodes (Springer Berlin Heidelberg New York, 2005) p. 253.

\bibitem{nelson1971theory}
D. F. Nelson, and M. Lax. Theory of the photoelastic interaction. Phys. Rev. B. \textbf{3}, 2778 (1971).

\bibitem{johnson2002perturbation}
S. G. Johnson et al. Perturbation theory for Maxwell's equations with shifting material boundaries. Phys. Rev E. \textbf{65}, 066611 (2002).

%\bibitem{Bart}
%B. Graczykowski, A. Gueddida, B. Djafari-Rouhani, H. J. Butt and G. Fytas. Brillouin light scattering under one-dimensional confinement: Symmetry and interference self-canceling. Phys. Rev. B \textbf{99}, 165431 (2019).

\end{thebibliography}
\end{document}